\documentclass[journal=ancac3,manuscript=letter,layout=twocolumn]{achemso}
\setkeys{acs}{maxauthors = 0}
\setkeys{acs}{articletitle = true}

\usepackage[version=3]{mhchem} % Formula subscripts using \ce{}
\usepackage{graphicx}% Include figure files
\usepackage{dcolumn}% Align table columns on decimal point
\usepackage{bm}% bold math
\usepackage[mathlines]{lineno}% Enable numbering of text and display math
\usepackage{amssymb}
\usepackage{amsmath}% bold math
\usepackage{soul,color}
\usepackage{dcolumn}% Align table columns on decimal point

\newcommand{\modR}[1]{{#1}}
\title{
       Degenerately Doped Transition Metal Dichalcogenides
       as Ohmic Homojunction Contacts to
       Transition Metal Dichalcogenide Semiconductors}%

\author{Zhibin Gao}
\affiliation{Physics and Astronomy Department,
             Michigan State University,
             East Lansing, Michigan 48824, USA}
\alsoaffiliation{Center for Phononics and Thermal Energy Science,
             School of Physics Science and Engineering,
             Tongji University, 200092 Shanghai, P. R. China}
\author{Zhixian Zhou}
\affiliation{Physics and Astronomy Department,
             Wayne State University,
             Detroit, Michigan 48201, USA}

\author{David Tom\'{a}nek}
\email{tomanek@pa.msu.edu} %
\affiliation{Physics and Astronomy Department,
             Michigan State University,
             East Lansing, Michigan 48824, USA}
\alsoaffiliation{Mandelstam Institute for Theoretical Physics
             and School of Physics,
             University of the Witwatersrand,
             2050 Johannesburg, South Africa}

\date{\today}

\begin{document}

%%%%%%%%%%%%%%%%%%%%%%%%%%%%%%%%%%%%%%%%%%%%%%%%%%%%%%%%%%%%%%%%%%%%%
%% The manuscript does not need to include \maketitle, which is
%% executed automatically.  The document should begin with an
%% abstract, if appropriate.  If one is given and should not be, the
%% contents will be gobbled.
%%%%%%%%%%%%%%%%%%%%%%%%%%%%%%%%%%%%%%%%%%%%%%%%%%%%%%%%%%%%%%%%%%%%%

\begin{abstract}
In search of an improved strategy to form low-resistance contacts
to MoS$_2$ and related semiconducting transition metal
dichalcogenides, we use {\em ab initio} density functional
electronic structure calculations in order to determine the
equilibrium geometry and electronic structure of MoO$_3$/MoS$_2$
and MoO$_2$/MoS$_2$ bilayers. Our results indicate that, besides a
rigid band shift associated with charge transfer, the presence of
molybdenum oxide modifies the electronic structure of MoS$_2$ very
little. We find that the charge transfer in the bilayer provides a
sufficient degree of hole doping to MoS$_2$, resulting in a highly
transparent contact region.
\end{abstract}

%\section*{Introduction}

One of the key challenges in 2D semiconductor physics is finding
better ways to inject charge carriers through transparent, ohmic
contacts~\cite{%
allain2015electrical,%
DT213,%
kang2014computational} %
in order to reduce power dissipation and increase carrier
mobility~\cite{%
lopez2013ultrasensitive,%
cui2015multi}. %
This problem does not occur in 3D semiconductors, where degenerate
doping, resulting in transparent contacts to metal electrodes, is
achieved by ion
implantation~\cite{%
yu1970electron}. That approach is not practical in atomically thin
2D layers, since bombardment by ions with a finite penetration
depth would not only implant, but also knock out atoms from the
channel.~\cite{fang2013degenerate} Several alternative approaches
have been explored in the past. The most common among these is
doping
% provided
by elemental substitution~\cite{%
yang2014chloride,%
suh2014doping,%
chuang2016low} %%
and %%
by surface charge transfer~\cite{%
fang2013degenerate,%
fang2012high,%
kiriya2014air}. %
%%or charge transfer from adjacent layers~\cite{%
%%cho2015phase,%
%%kappera2014phase,%
%%das2012high}.
Doping the entire structure does improve charge injection in the
contact region, but also turns the channel metallic, reducing its
switching capability~\cite{xzheng17}. Significant local doping by
surface charge transfer may also be achieved using adsorbed
molecules,
% substantial,
but typically suffers from lack of chemical and thermal stability.
Substitutional doping yields a stable structure,
%% prohibits selectively doping  %%
but cannot be strictly limited to the contact regions only. It
generally causes structural changes and introduces scattering
centers in the 2D channel, thus degrading its transport
properties. Direct local contact between a 2D semiconductor and a
metal with a favorable work function appears attractive, but
typically results in a strong hybridization at the interface,
causing the formation of mid-gap states and Fermi level
pinning~\cite{%
{DT243},
kang2014computational,%
chen2013tuning,%
gong2014unusual,%
movva2015high}. One way to %%
%% avoid
reduce %%
these negative side effects of a metal contact is to insert %%
%% a 2D monolayer
an ultra-thin insulating layer such as h-BN or
%% graphene
MgO %%
in-between the semiconducting channel and the metal~\cite{%
%%chuang2014high,%
%%liu2015toward,%
{chen2013control},%
{wang2016high},%
{cui2017low}}. %
This approach favorably modifies the work function of the metal
while reducing the net charge transfer between the contact metal
and the channel, suppresses the formation of interlayer gap states
and Fermi level pinning. The drawback of this approach is widening
the vertical tunnel barrier between the contact metal and the
channel. Whereas formation of mid-gap states can be avoided when
using
%% a substitutionally doped 2D semiconductor
substitutionally heavily doped or metallic 2D transition metal
dichalcogenides
%% such as NbSe$_2$ and Nb$_{0.005}$W$_{0.995}$Se$_2$ %%
in van der Waals contact with the
channel\cite{{chuang2016low},{DT259}}, the vertical tunnel barrier
can not be eliminated. An unusual way to optimize contacts to 2D
semiconductors, which does not involve charge transfer to the
channel, is by phase
engineering~\cite{%
kappera2014phase,%
cho2015phase,%
yang2017structural}. %
In this approach, an in-layer homo-junction is formed when locally
converting the semiconducting 2H phase of the channel material
such as MoS$_2$ to to a metastable, metallic 1T phase. This
approach is, however, restricted to selected materials and the
phase stability is limited. All of the above strategies have
significant drawbacks,
% The drawbacks of all the above strategies are significant,
which make continuing search for suitable alternatives highly
desirable.

%===========< FIGURE 1 >=========================================
\begin{figure}[t!]
\includegraphics[width=0.85\columnwidth]{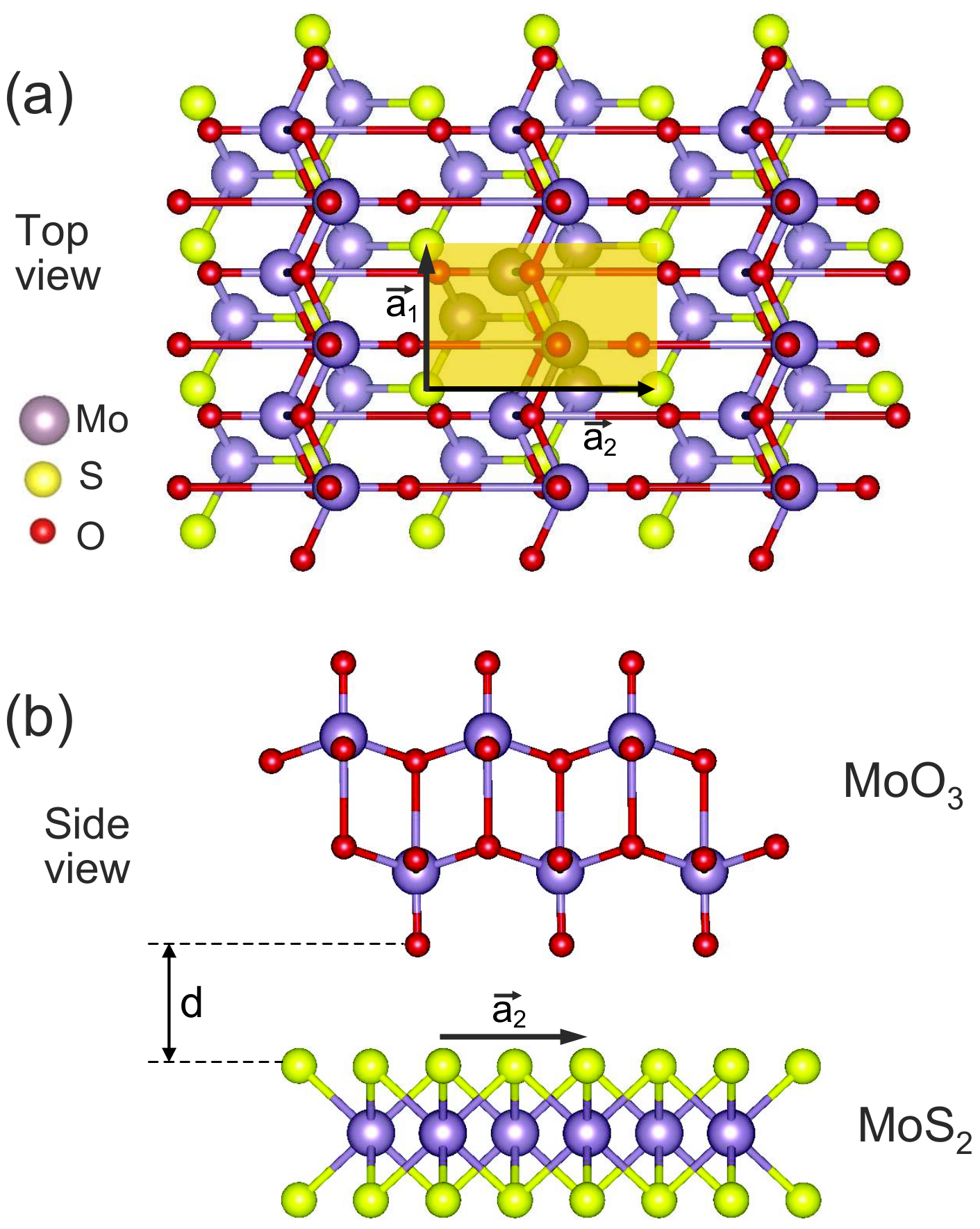}
\caption{Ball-and-stick model of the MoO$_3$/MoS$_2$ bilayer in
(a) top view and (b) side view. The unit cell is indicated by
yellow shading in (a). $\vec{a}_1$ and $\vec{a}_2$ are the lattice
vectors spanning the 2D lattice and $d$ is the inter-layer
distance. \label{fig1} }
\end{figure}
%===========< FIGURE 1 >=========================================

%\section*{Executive Summary}

\modR{%
We propose a previously unexplored approach to achieve degenerate
hole doping in a local region of a semiconducting MoS$_2$ channel
that is placed in direct contact with electronegative materials
such as MoO$_3$ or MoO$_2$ using van der Waals assembly. %
}%
%\modG{%
To prove the viability of this approach, we perform {\em ab
initio} density functional calculations of the vertical
heterostructure using density functional theory (DFT), which is
known to describe
charge distribution and doping correctly. %
%}%
We find that the charge transfer between the doping layer and the
channel is significant and results in degenerate doping of the
MoS$_2$ channel by the adjacent molybdenum oxide layer. With only
negligible hybridization at the bilayer interface and its negative
side effects, the dominant electronic structure change is a net
charge transfer giving rise to an interface dipole, which causes a
rigid shift~\cite{DT259} of the MoS$_2$ bands in the contact
region and turns it metallic. In this case, no vertical tunnel
barrier will form when the heavily doped and highly conducting
contact region of the channel material is contacted by a metal on
the side opposite to molybdenum oxide.

%\section*{Extended Introduction to the State of the Art}

In common Si-based 3D semiconductor devices, charge transport is
dominated by the intrinsic properties of the channel. Contacts do
not play a significant role, since the contact region is
degenerately doped by ion implantation, which allows for
barrier-free charge injection from a metal
electrode.~\cite{yu1970electron} The device characteristics change
significantly in low-dimensional semiconductors, where contacts
play the dominant role due to significant Schottky barriers and an
associated wide depletion region caused by insufficient screening.
The key role of Schottky barriers in the contact region has been
recognized early in 1D semiconducting carbon nanotubes with metal
contacts,~\cite{SchottkyCNT02} with similar behavior extending to
2D semiconductors with metal contacts.\cite{yi2015study} %%
%%~\cite{yu2016carrier}
There, the contact resistance can be reduced, albeit not
eliminated, by selecting contact metals with a work function
aligned with the valence band maximum (VBM) or conduction band
minimum (CBM), depending on the type of carriers to be
injected.~\cite{{DT243},{farmanbar2016ohmic}}

Our alternative approach is to to create a homo-junction between a
doped and a pristine segment of a 2D semiconductor such as
MoS$_2$. Key to this approach is the use of van der Waals assembly
to contact the channel by a highly electronegative material that
locally provides degenerate hole doping. The material of choice in
this study is MoO$_3$ with a very high work
function~\cite{greiner2012} of $6.7$~eV. This material has emerged
as a promising surface acceptor material for a wide variety of
systems including diamond thin films, graphene,
and transition metal dichalcogenides (TMDs)~%
\cite{%
xie2011electrical,%
russell2013surface,%
chuang2014mos2,%
xia2016surface}. %
Because the CBM of MoO$_3$ lies below the VBM of most
semiconductors including TMDs, molybdenum oxide is a viable
candidate to provide degenerate doping to 2D semiconductors.

%==================================================================

\section*{Results}

\subsection*{MoS$_2$ in contact with MoO$_3$}

The structure of isolated MoO$_3$ and MoS$_2$ monolayers, as
well as that of the bilayer, is shown in Figure~\ref{fig1}. %
The Bravais lattice of MoO$_3$ has a rectangular unit cell
containing 2 Mo and 6 O atoms with
$a_1=3.71$~{\AA} and %
$a_2=3.95$~{\AA}. %
% MoO3 a1=3.706A, a2=3.9549A
The calculated cohesive energy per formula unit of a free-standing
MoO$_3$ monolayer is $E_{coh}=24.62$~eV with respect to isolated
atoms.
% Q1. 1 Mo + 3 O atoms --> free-standing MoO3 monolayer (relaxed):
%     how much energy gain? $E_{coh}=24.619~eV
%     free-standing MoO3 monolayer (relaxed) = -32.89140121 eV;
%     Mo atom: -3.68736057 eV; O atom: -1.52822916 eV

MoS$_2$ forms a triangular lattice with the lattice constant
$a=3.18$~{\AA}. The corresponding lattice constants of a
conventional rectangular unit cell containing 2 Mo and 4 S atoms,
useful when discussing epitaxy with MoO$_3$, are %
$a_1=3.18$~{\AA} and $a_2=5.51$~{\AA}. %
% MoS2: a=3.18A triangular or a1=3.18A, a2=5.51A
The calculated cohesive energy per formula unit of free-standing
MoS$_2$ is $E_{coh}=16.37$~eV with respect to isolated atoms.
% Q3. 1 Mo + 2 S atoms --> free-standing MoS2 monolayer (relaxed):
%     how much energy gain? $E_{coh}=16.36771994~eV
%     free-standing MoS2 monolayer (relaxed) = -21.79870791 eV;
%     Mo atom: -3.68736057 eV; S atom: -0.87181370 eV

%===========< FIGURE 2 >=========================================
\begin{figure*}[t]
\includegraphics[width=1.6\columnwidth]{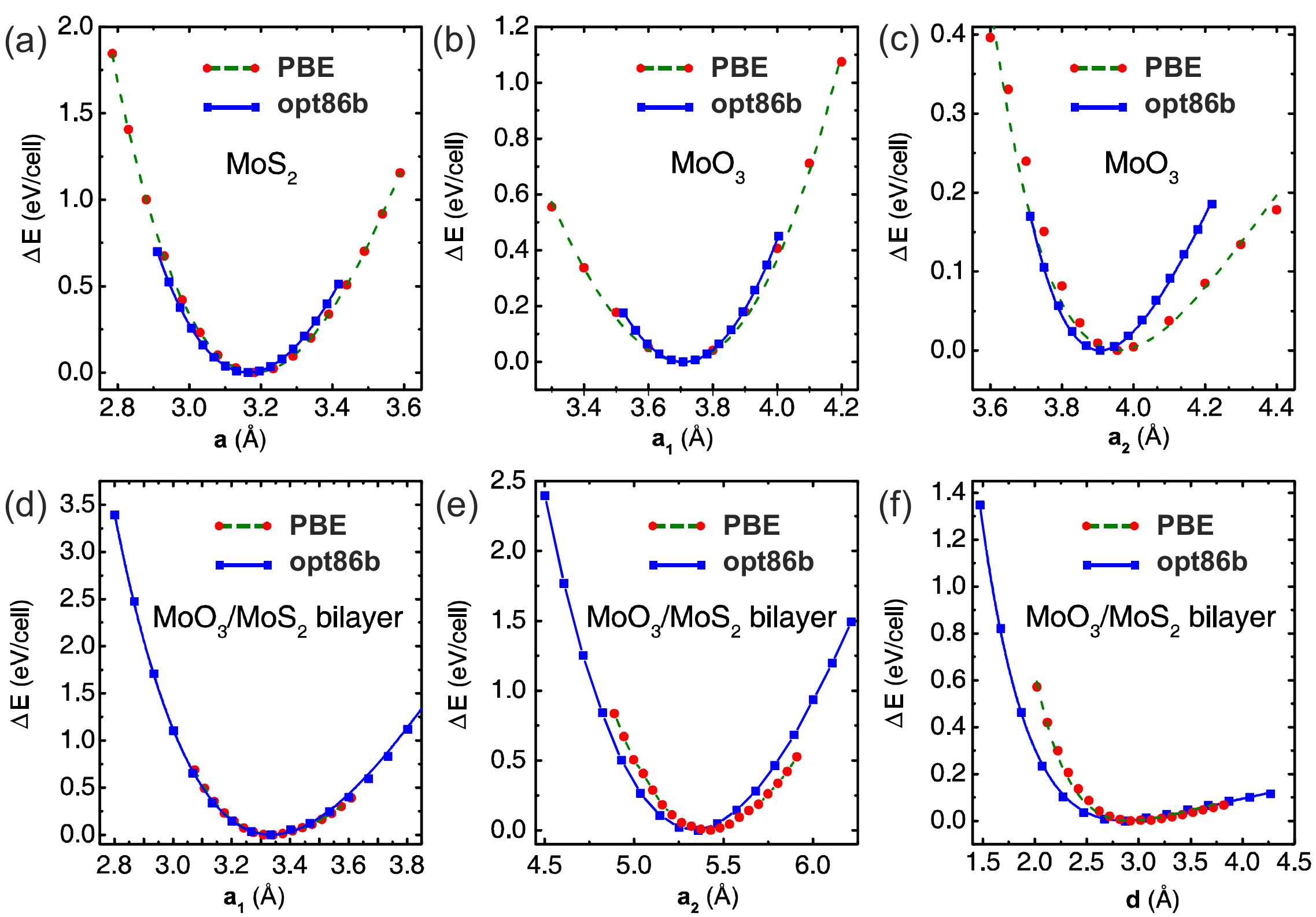}
\caption{Energy change ${\Delta}E$ per unit cell as a function of
the lattice parameters. Monolayer results are presented for (a)
MoS$_2$ with threefold symmetry and lattice constant $a$ and
(b),(c) MoO$_3$ with lattice constants $a_1$ and
$a_2$. %
% per 3 atoms
The unit cells we consider contain 3 atoms in (a)
% per 8 atoms
and 8 atoms in (b),(c). %
MoO$_3$/MoS$_2$ bilayer results representing 14-atom unit cells
show ${\Delta}E$ as a function of (d) $a_1$, (e) $a_2$, and (f)
the interlayer distance $d$. The dashed lines in (a-d) and (f) are
fits by the Morse potential.
% ${\Delta}E=D(1-e^{{\xi}(a-a_{0})})$
% Energy in (d-e): per 14 atoms in supercell A
The dashed line in (e) is a guide to the eye. %
%
%\modG{ %
The legend in the panels specifies the symbols and line types for
results based on DFT-PBE and also the DFT-optB86b-vdW functional,
which specifically considers the van der Waals interaction. %
%}%
\label{fig2} }
\end{figure*}
%===========< FIGURE 2 >=========================================

%===========< FIGURE 3 >=========================================
\begin{figure*}[t]
\includegraphics[width=1.8\columnwidth]{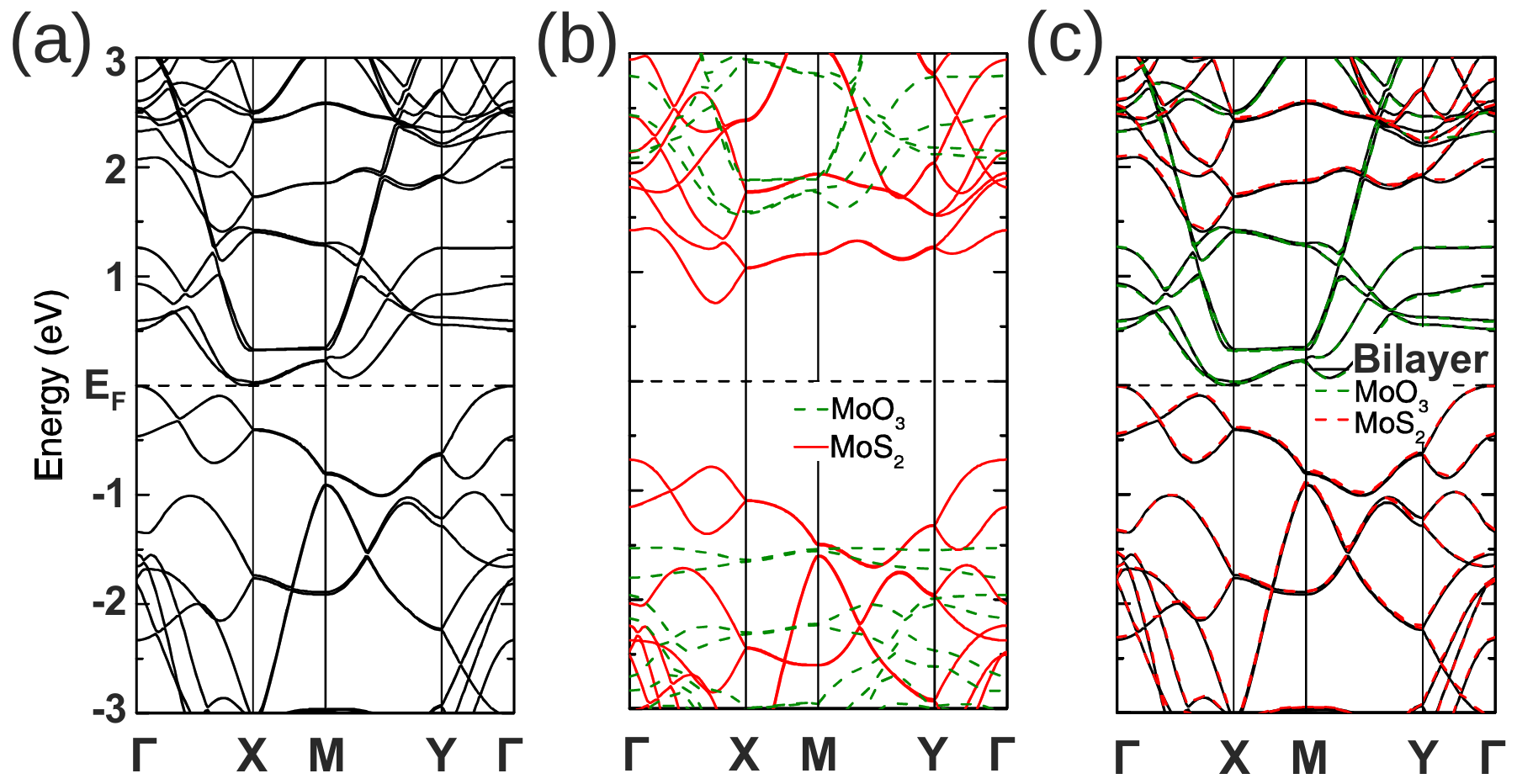}
\caption{(a) Electronic band structure of a MoO$_3$/MoS$_2$
bilayer
\modR{%
with supercell A.} %
(b) Superposition of the electronic band structure of isolated
monolayers of MoS$_2$ (solid red lines) and MoO$_3$
(dashed green lines). %
(c) Superposition of the MoS$_2$ bands in panel (b), shifted
rigidly up by $0.711$~eV, and the MoO$_3$ bands, shifted rigidly
down by $1.523 $~eV. The combined band structure in (c) is
superposed to that of the MoO$_3$/MoS$_2$ bilayer in panel (a). %
%\modG{%
%Results in (a)-(c) are based on DFT-PBE. Panel (d) is the
%counterpart of (c) obtained using DFT-HSE06. %
%}%
\label{fig3} }
\end{figure*}
%===========< FIGURE 3 >=========================================

%===========< FIGURE 4 >=========================================
\begin{figure*}[h!]
\includegraphics[width=1.6\columnwidth]{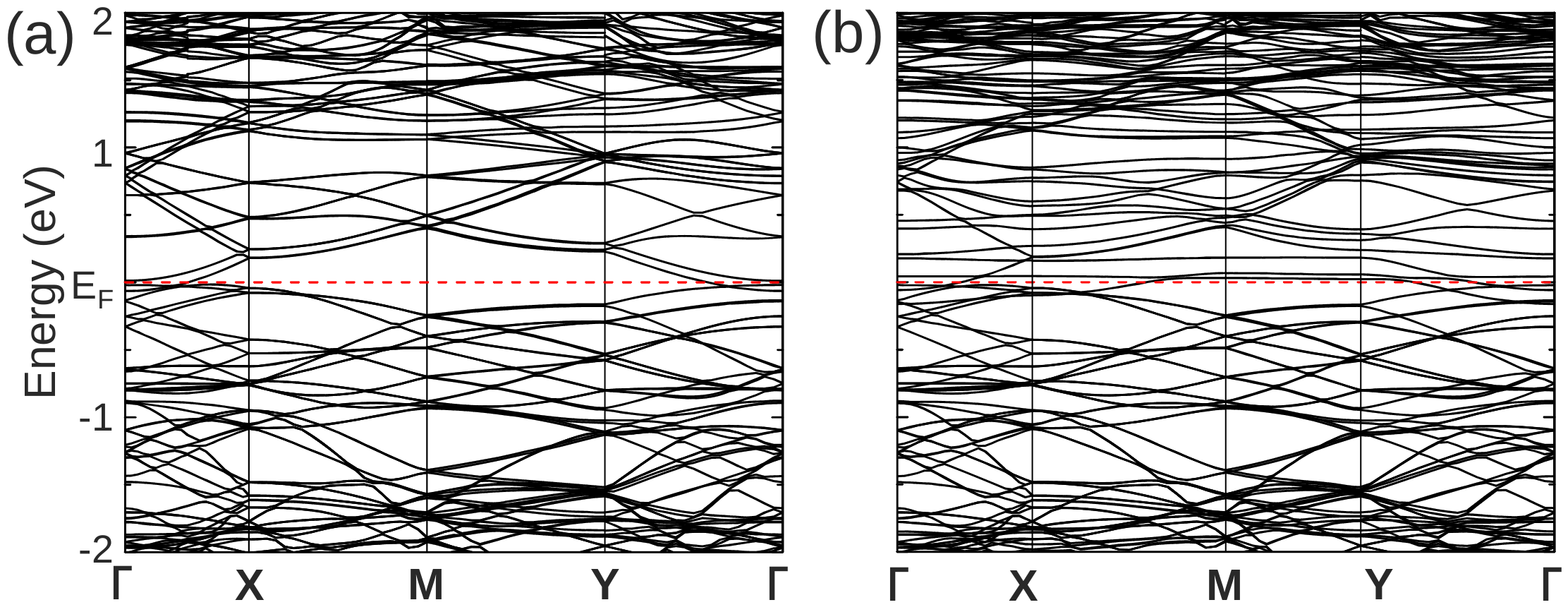}
\caption{%
\modR{%
Electronic band structure of an MoO$_{3-x}$/MoS$_2$ bilayer based
on DFT-PBE and represented by supercell B containing 156 atomic
sites in total. Results for a defect-free system ($x=0$) (a) are
compared to those for a defective system ($x>0$)
with one oxygen vacancy in each unit cell (b). %
The Fermi level is indicated by the red dashed line.
}%
\label{fig4} }
\end{figure*}
%===========< FIGURE 4 >=========================================

%===========< FIGURE 5 >=========================================
\begin{figure*}[t]
\includegraphics[width=1.5\columnwidth]{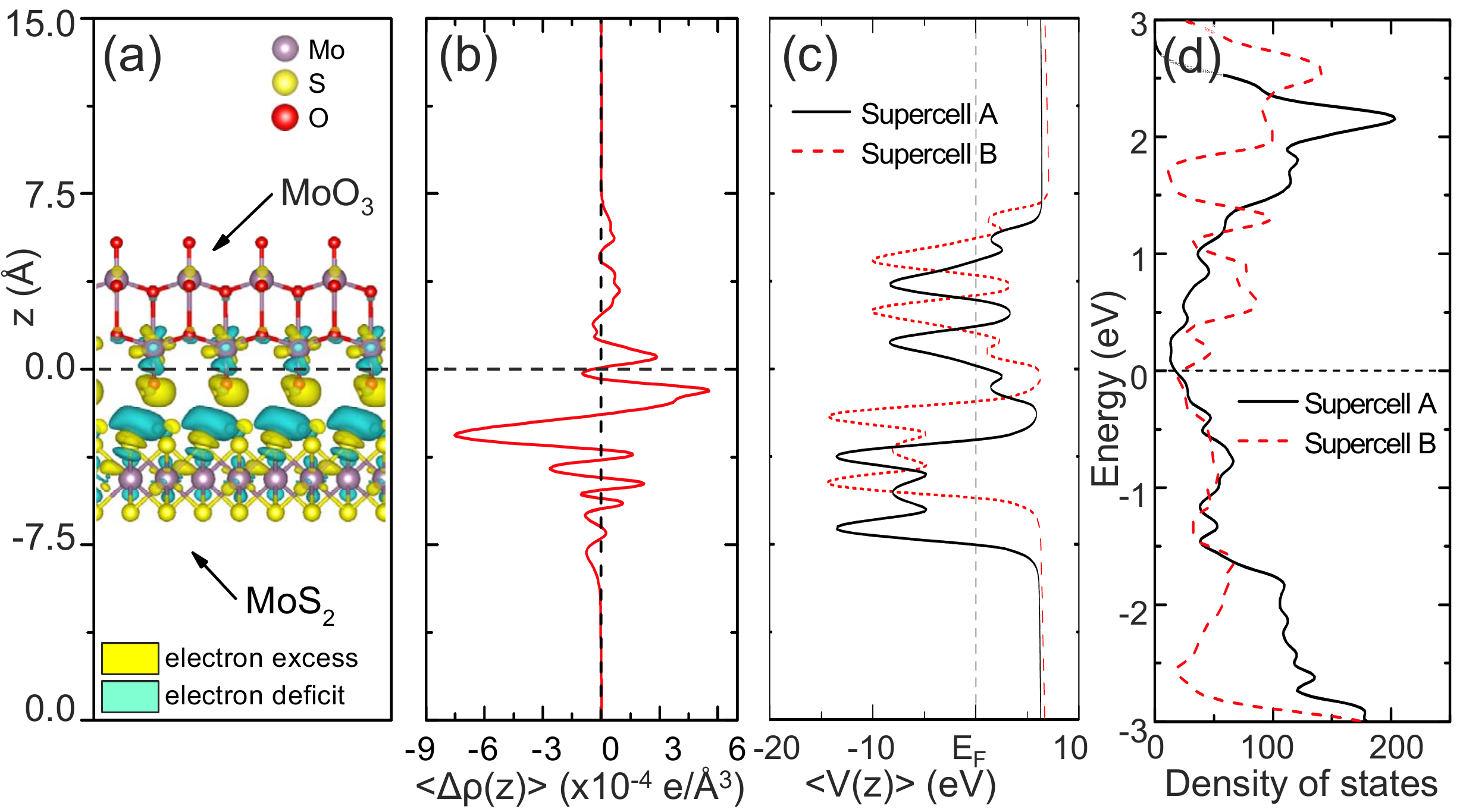}
\caption{Electronic structure changes associated with assembling
the MoO$_3$/MoS$_2$ bilayer from isolated monolayers. %
(a) Charge density difference
${\Delta}\rho=\rho({\rm{MoO}}_3/{\rm{MoS}}_2) %
-\rho({\rm{MoO}}_3) %
-\rho({\rm{MoS}}_2)$. %
${\Delta}\rho$ is shown by isosurfaces bounding regions of
electron excess at $+1.0\times10^{-3}~\text{e}$/{\AA}$^3$ (yellow)
and electron deficiency at
$-1.0\times10^{-3}~\text{e}$/{\AA}$^3$ (blue). %
(b) $\langle{\Delta}\rho(z)\rangle$ averaged across the $x-y$
plane of the layers. The black solid reference line is a guide to
the eye. $z$ indicates the position of the plane. %
(c) Electrostatic potential $<V(z)>$ in the bilayer, averaged
across the x-y plane, with $z$ denoting the position of the plane.
% The lattice constant along z direction is $30$~{\AA}.
(d) Density of states (DOS) of the bilayer, convoluted by a
Gaussian with a full-width at half-maximum of $0.1$~eV. %
(c) and (d) display results for supercell A (solid black line) and
supercell B (dashed red line), defined in the text. %
%\modR{%
%Results in (a)-(c) and supercell B in (d) are based on DFT-PBE.
%Supercell A in panel (d) is the counterpart DOS obtained using
%DFT-HSE06.}
% Supercell A is shown in Figure 1(a)
% The lattice constant along z direction is $30$~{\AA}.
\label{fig5} }
\end{figure*}
%===========< FIGURE 5 >=========================================

Since the computational approach to investigate infinite
structures requires a periodic array of identical unit cells and
since the MoO$_3$ and the MoS$_2$ layers are incommensurate, we
need to enforce commensurability by subjecting the individual
layers to in-layer stain at some energy cost. %
%\modG{%
To obtain quantitative understanding of the effect of strain on
our results, we distinguished the smaller supercell A containing
4~Mo, 6~O, and 4~S atoms, shown in Figure~\ref{fig1}(a), from the
larger supercell B that contains 44~Mo, 72~O, and 40~S atoms.
Both supercells are shown in the Supporting Information. %
% Fig. S1
%}%
\modR{%
The primary reason to consider the smaller supercell A, which is
highly strained, is to obtain an intuitive understanding of the
Physics. The larger supercell B is less strained and expected to
provide a better quantitative agreement with the real system. %
}%

The strain energy in stretched and compressed MoO$_3$ and MoS$_2$
monolayers as well as in the bilayer is shown in
Figure~\ref{fig2}.
% Formula unit: 2 Mo, 3 O, 2 S
% Supercell A: 4 Mo, 6 O, 4 S, total 14 atoms
% Supercell B: 44 Mo, 72 O, 40 S atoms, total 156 atoms
The deformation energies ${\Delta}E$, shown as a function of the
lattice constants in Figure~\ref{fig2}(a)-\ref{fig2}(e), allow to
judge the in-plane compressibility of the MoO$_3$/MoS$_2$ bilayer
and its components. The inter-layer interaction energy in the
bilayer system with supercell A, shown in Figure~\ref{fig2}(f),
has been fitted by the Morse potential
${\Delta}E=E_0\left[1-\exp{\left({\xi}[d-d_{0}]\right)}\right]$. %

\modR{ %
We find results based on DFT-PBE to be in very good agreement with
those based on the DFT-optB86b-vdW functional. This indicates that
the role of van der Waals interactions, which are specifically
addressed in the latter functional, is only secondary. Since our
results for both supercells indicate a significant charge
redistribution within the vertical heterojunction, the role of the
Coulomb attraction between adjacent MoO$_3$ and MoS$_2$ layers is
more significant. We note that Coulomb interaction is
described adequately by the DFT formalism. %
}%

The relaxed geometry of the the bilayer structure with the
rectangular supercell A is characterized by the lattice
constants $a_1=3.34$~{\AA}, $a_2=5.42$~{\AA}, %
% MoO3/MoS2 Supercell A: a1=3.34175A a2 = 5.42A
and the unit cell area $A_A=18.11$~{\AA$^2$}. We note that the
optimum values of $a_1$ and $a_2$ in the bilayer lie in-between
the values of the isolated monolayers. In the MoO$_3$ layer within
the relaxed bilayer with supercell A, $a_1$ is compressed by 9.8\%
and $a_2$ is stretched by 37.0\% at the energy cost of
$5.5$~eV/nm$^2$. In the MoS$_2$ layer in the same geometry, $a_1$
is stretched by 5.1\% and $a_2$ is compressed by 1.6\% at the
energy cost of $0.9$~eV/nm$^2$. The attractive interaction energy
between the strained layers amounts to $0.1$~eV/nm$^2$.

The relaxed geometry of the the bilayer structure with the larger
rectangular supercell B is characterized by the lattice constants
$a_1=15.87$~{\AA} and $a_2=11.07$~{\AA} and the unit cell area
$A_B=175.72$~{\AA$^2$}. We note again that the optimum $a_1$ and
$a_2$ values in the bilayer are in-between the values of the
isolated monolayers. The relative deformations and the deformation
energies are much smaller in the larger supercell B. In the
MoO$_3$ layer within the relaxed bilayer with supercell B, $a_1$
is stretched by 7.0\% and $a_2$ is compressed by 6.7\% at the
energy cost of $0.3$~eV/nm$^2$. In the MoS$_2$ layer within the
same relaxed bilayer geometry, $a_1$ is compressed by 0.2\% and
$a_2$ is stretched by 0.5\% at the energy cost of $0.6$~eV/nm$^2$.
The attractive interaction energy between the strained layers in
the bilayer amounts to $0.7$~eV/nm$^2$, nearly compensating for
the deformation energy of the individual layers.

%==================================================================

\modR{%
To obtain a basic understanding of what is happening in the system
electronically, we first consider the smaller supercell A of the
MoO$_3$/MoS$_2$ bilayer. The electronic band structure based on
DFT-PBE Kohn-Sham eigenvalues, shown in Figure~\ref{fig3}(a),
suggests that the system is semimetallic. A superposition of
electronic band structures of isolated MoO$_3$ and MoS$_2$
monolayers, constrained to their respective structure in the
bilayer, is shown in Figure~\ref{fig3}(b), with the Fermi levels
aligned at mid-gap. In view of the fact that fundamental gaps are
typically underestimated in the Kohn-Sham DFT-PBE spectrum, we
expect that the MoO$_3$/MoS$_2$ bilayer with supercell A, as well
as the MoO$_3$ and MoS$_2$ monolayers, should all be
semiconducting. This expectation is confirmed in the corresponding
band structure and density of states obtained using the hybrid
DFT-HSE06 exchange-correlation functional, which is presented in
the Supporting Information.
% Fig. S2(b), Fig. S4(b)
Kohn-Sham spectra based on DFT-HSE06 typically have wider
fundamental band gaps that agree better with
observed electronic spectra than spectra based on DFT-PBE.%
}%

To inspect the level of hybridization at the interface, we
superposed the DFT-PBE band structure of the bilayer with the
bands of an isolated MoS$_2$ monolayer, shifted rigidly up by
$0.711$~eV, and an MoO$_3$ monolayer, shifted rigidly down by
$1.523 $~eV. Comparing the band superposition with the bilayer
band structure in Figure~\ref{fig3}(c), we find only a minimum
degree of hybridization between the individual layers.
% Figures S2(b) and S4(b): HSE06, same structure as in PBE.
% HSE06 band gap is Eg = 0.5644 eV wide.
% Superposition of MoS2 bands shifted rigidly up by 0.6597 eV
% and MoO3 bands shifted rigidly down by 2.4608 eV.
\modR{%
The same small degree of hybridization also occurs in DFT-HSE06
calculations. Similar to DFT-PBE results, the DFT-HSE06 band
structure of the bilayer is represented well by a superposition of
MoS$_2$ monolayer bands, shifted rigidly up by $0.660$~eV, and
MoO$_3$ monolayer bands, shifted rigidly down by $2.461$~eV.
}%

\modR{%
In truth, the above semiconducting behavior is not intrinsic to
the MoO$_3$/MoS$_2$ bilayer, but rather linked to using the small
supercell A. DFT-PBE band structure results for the bilayer
represented by the much larger and more adequate supercell B,
presented in Fig.~\ref{fig4}(a), indicate a much higher degree of
metallicity at the Fermi level. In realistic MoO$_{3-x}$ systems,
there are non-stoichiometric composition variations with oxygen
vacancies~\cite{{chuang2014mos2},{balendhran2013enhanced}}. As
shown in Fig.~\ref{fig4}(b), such oxygen vacancies always cause
partly filled bands associated with metallic behavior.
}%

The charge redistribution in the bilayer is shown in
Figure~\ref{fig5}. The net charge transferred from
the MoS$_2$ to the MoO$_3$ layer is %
${\Delta}\rho_{2D}=2.0{\times}10^{13}$~e/cm$^2$ in the bilayer
%${\Delta}\rho_{2D}=1.973{\times}10^{13}$~e/cm$^2$
with the smaller supercell A and %
${\Delta}\rho_{2D}=4.5{\times}10^{13}$~e/cm$^2$ %
%${\Delta}\rho_{2D}=4.4913{\times}10^{13}$~e/cm$^2$
in the bilayer with the larger supercell B.
% As we could expect, the charge transfer is larger in the
% system with larger supercell B with lower strain.
When divided by the thickness $t{\approx}0.4$~nm of the MoS$_2$
channel, the charge transfer density ranges from
${\Delta}\rho=5{\times}10^{20}-1{\times}10^{21}$~e/cm$^3$. This is
high enough and considered degenerate doping, since $E_F$ has been
moved into the valence band. Having achieved degenerate doping in
the contact region of the bilayer, the tunnel barrier to a metal
contact at the side opposite to the doping layer should also be
negligibly small and of no consequence.

%==================================================================

\subsection*{MoS$_2$ in contact with MoO$_2$}

%\modG{%
As mentioned before, experimental studies report
non-stoichiometric variations of MoO$_{3-x}$ with oxygen vacancies
and amorphous
structure~\cite{{chuang2014mos2},{balendhran2013enhanced}}, which
cannot be modelled as periodic structures. As a step in the
direction of molybdenum oxide with a lower oxygen concentration,
we consider instead the MoO$_2$ stoichiometry.
%}%
MoO$_2$ has been discussed theoretically and found to form a
stable honeycomb lattice~\cite{ataca2012stable}, different from
the MoO$_3$ lattice with a rectangular unit cell. Since MoO$_2$
shares the same sixfold symmetry with MoS$_2$, we have considered
this system as a potential alternative to MoO$_3$ in hole-doping
MoS$_2$.

%===========< FIGURE 6 >=========================================
\begin{figure}
\includegraphics[width=0.8\columnwidth]{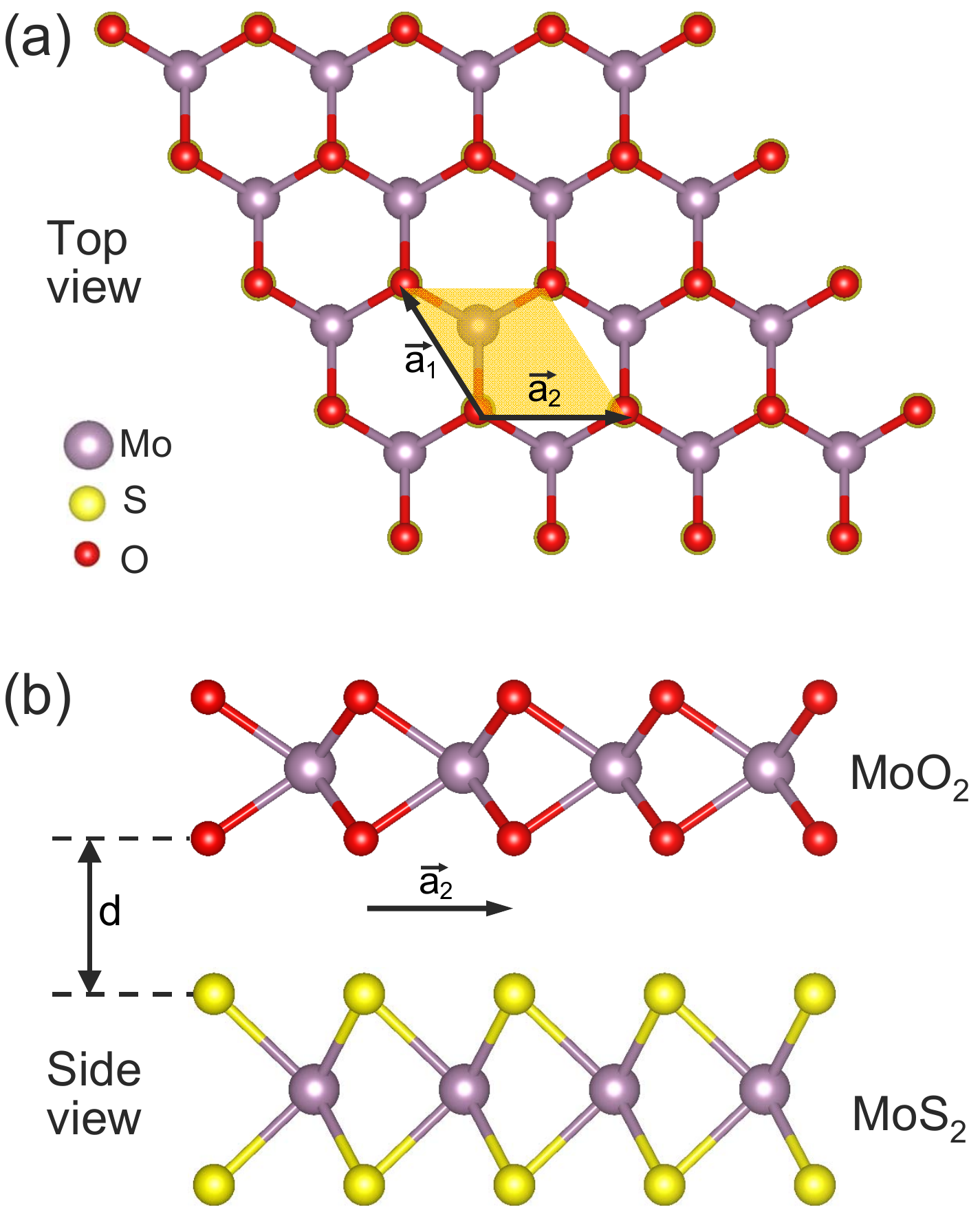}
\caption{Ball-and-stick model of the MoO$_2$/MoS$_2$ bilayer in
(a) top view and (b) side view. The unit cell is indicated by
yellow shading in (a). $\vec{a}_1$ and $\vec{a}_2$ are the lattice
vectors spanning the 2D lattice and $d$ is the inter-layer
distance. %
\label{fig6} }
\end{figure}
%===========< FIGURE 6 >=========================================

The structure of the isolated MoO$_2$ and MoS$_2$ monolayers, as
well as the bilayer structure, are shown in Figure~\ref{fig6}. %
Unlike MoO$_3$, MoO$_2$ forms a honeycomb lattice with the lattice
constant $a=2.83$~{\AA}. %
% a=2.83006 A
The calculated cohesive energy per formula unit of a free-standing
MoO$_2$ monolayer is $E_{coh}=19.65$~eV with respect to isolated
atoms.
% Q2. 1 Mo + 2 O atoms --> free-standing MoO2 monolayer (relaxed):
%     how much energy gain? $E_{coh}=19.64989839$~eV
%     free-standing MoO2 monolayer (relaxed) = -26.39371728 eV;
%     Mo atom: -3.68736057 eV; O atom: -1.52822916 eV

We describe the MoO$_2$/MoS$_2$ bilayer by an epitaxial triangular
Bravais lattice with a basis and one formula unit per unit cell.
The optimum lattice constant is
$a=|\vec{a_1}|=|\vec{a_2}|=2.99$~{\AA} %
% a=2.985A
and the unit cell area is $A=7.72$~{\AA$^2$}. We note that the
optimum value of $a$ in the bilayer are in-between the values of
the isolated monolayers. In the MoO$_2$ layer, $a$ is stretched by
5.5\% at the energy cost of $5.2$~eV/nm$^2$. In the MoS$_2$ layer,
$a$ is compressed by 6.1\% at the energy cost of $4.5$~eV/nm$^2$.
The attractive interaction energy between the strained layers
amounts to $0.1$~eV/nm$^2$.

%===========< FIGURE 7 >=========================================
\begin{figure*}[t]
\includegraphics[width=1.6\columnwidth]{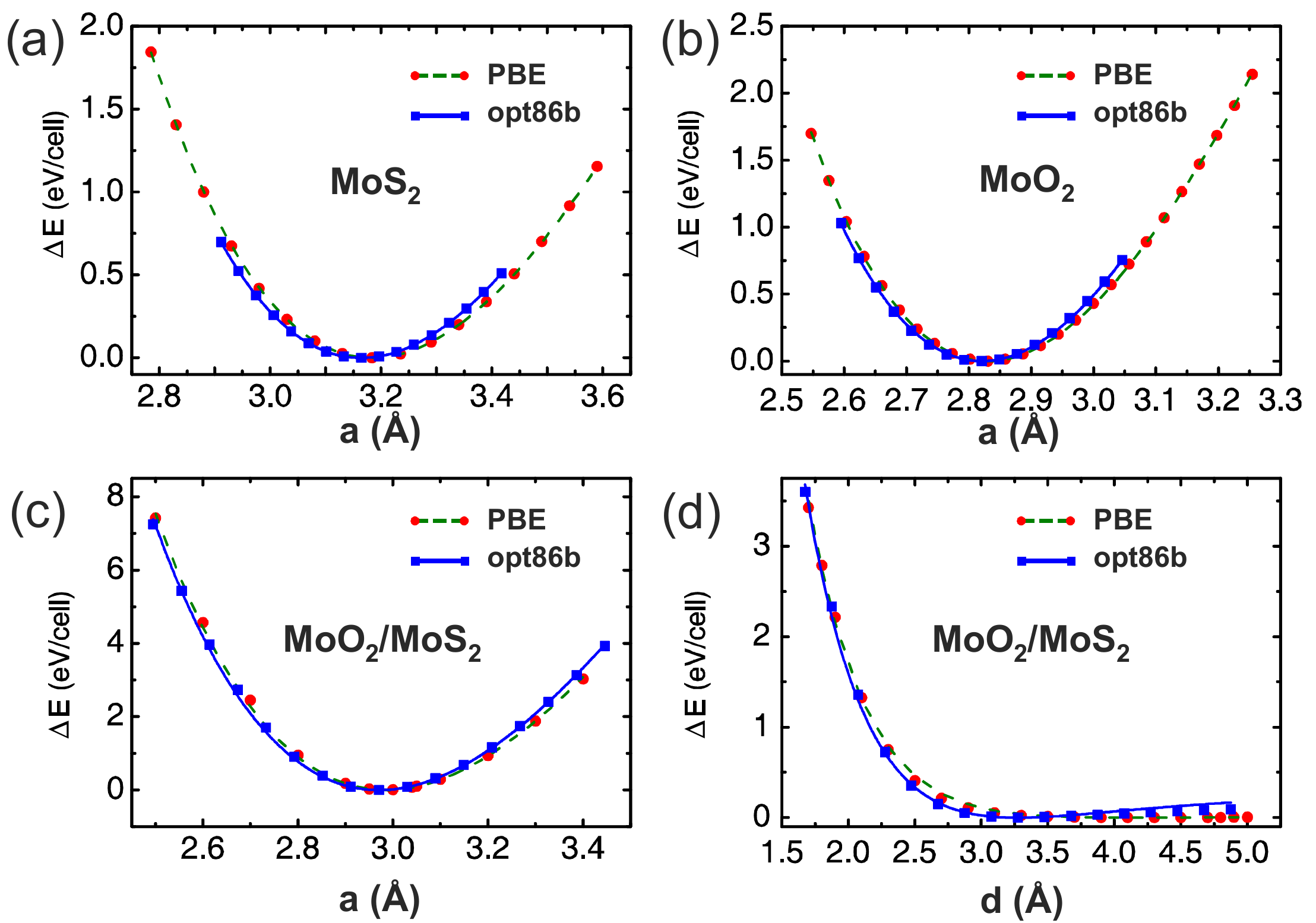}
\caption{Energy change ${\Delta}E$ per unit cell as a function of
the lattice constant $a$ in monolayers of %
(a) MoS$_2$ and %
(b) MoO$_2$. %
$\Delta$E per unit cell in MoO$_2$/MoS$_2$ bilayers as a function
of (c) the lattice constant $a$ and (f) the interlayer
distance $d$. The dashed lines are fits by the Morse potential. %
%
%\modG{ %
The legend in the panels specifies the symbols and line types for
results based on DFT-PBE and for those based on the
DFT-optB86b-vdW functional, which specifically considers the van
der Waals interaction. %
%}%
\label{fig7} }
\end{figure*}
%===========< FIGURE 7 >=========================================

%===========< FIGURE 8 >=========================================
\begin{figure*}[h!]
\includegraphics[width=1.5\columnwidth]{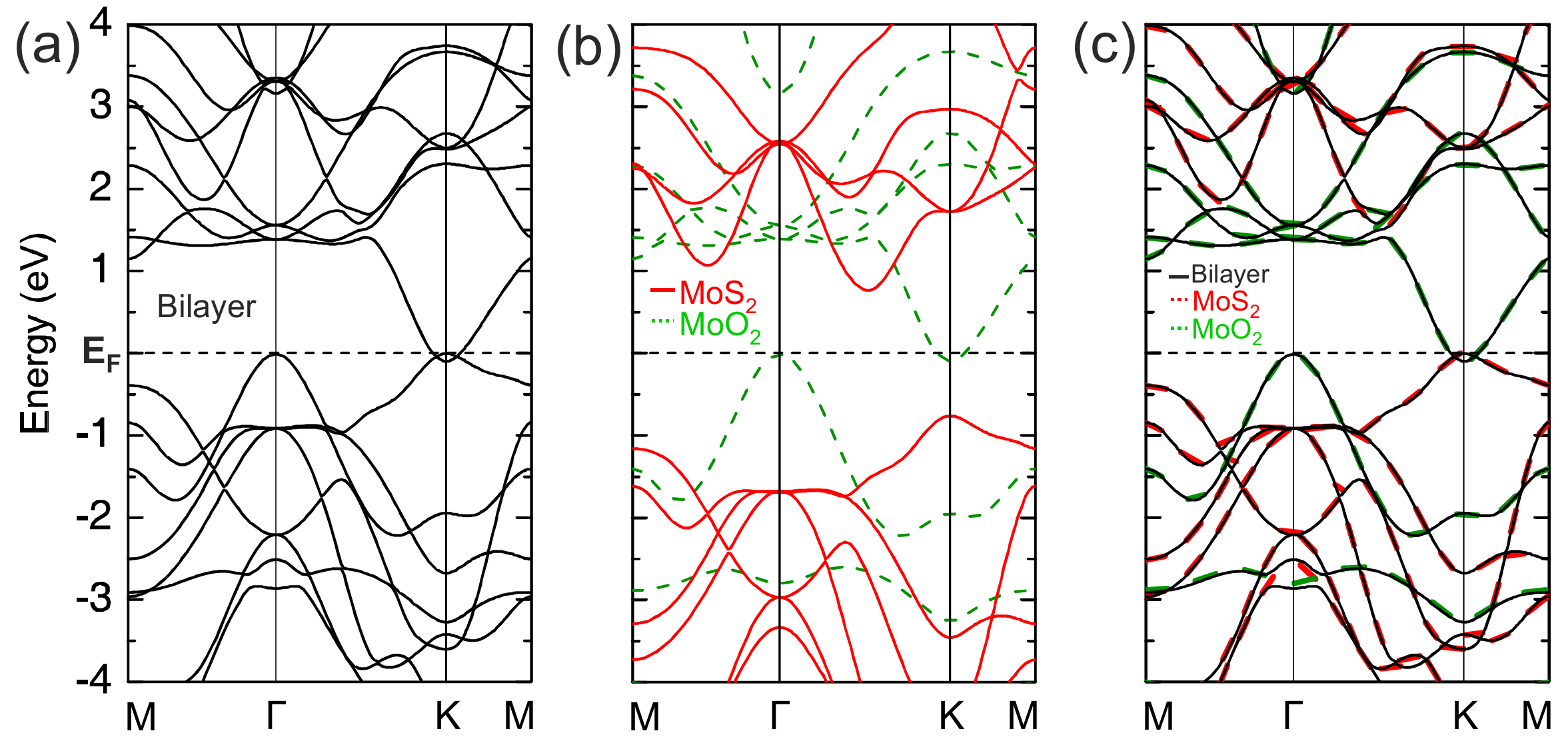}
\caption{(a) Electronic band structure of a MoO$_2$/MoS$_2$
bilayer. %
(b) Superposition of the electronic band structure of isolated
monolayers of MoS$_2$ (solid red lines) and MoO$_2$
(dashed green lines). %
(c) Superposition of the MoS$_2$ bands in panel (b), shifted
rigidly up by $0.768$~eV, and the unshifted MoO$_2$ bands. The
combined band structure in (c) is superposed to that of
the MoO$_2$/MoS$_2$ bilayer in panel (a). %
%\modR{%
%Results in (a)-(c) are based on DFT-PBE. Panel (d) is the
%counterpart of (c) obtained using DFT-HSE06. %
%}%
\label{fig8} }
\end{figure*}
%===========< FIGURE 8 >=========================================

Similar to MoO$_3$ interacting with MoS$_2$, also interacting
monolayers of MoO$_2$ and MoS$_2$ need to be strained to form a
commensurate structure. The unit cell of the bilayer, shown in
Figure~\ref{fig6}(b), contains 2~Mo, 2~S, and 2~O atoms. The
strain energy in stretched and compressed MoO$_2$ and MoS$_2$
monolayers as well as the bilayer is shown in Figure~\ref{fig7}. %
%\modG{ %
Also in this system, we find results based on DFT-PBE to be in
very good agreement with those based on the DFT-optB86b-vdW
functional. This indicates that van der Waals interactions, which
are specifically considered in the latter functional, play only a
secondary role.
%}%

% Supercell: 2 Mo 2 O, 2 S, total 6 atoms
%
% Q10. Isolated, strained MoO2 +
%   isolated, strained MoS2 -->
%   MoO2/MoS2 bilayer: how much energy gain?
%   How many Mo atoms, how many O atoms, how many S atoms?
% There are 2 Mo atoms, 2 O atoms and 2 S atoms: 6 atoms
% MoO2/MoS2 a=2.985A, mismatch: 6.1% wrt MoS2 (hexagonal lattice)
% Isolated, strained MoO2 (supercell A) = -26.03202672 eV
% isolated, strained MoS2 (supercell A) = -21.40133632 eV MoO2/MoS2
% bilayer with supercell A = -47.44338113 eV
% Energy gain = 0.01001809 eV
%   This is interaction energy.
%
% Q11. Isolated, strained MoO2 -->
%      isolated, relaxed MoO2: how much energy gain?
% MoO2/MoS2 a=2.985A, mismatch: 5.5% wrt MoSO (hexagonal lattice)
% Isolated, strained MoO2 = -26.03202672 eV
% isolated, relaxed MoO2 = -26.39371728 eV
% Energy gain = 0.36169056 eV per formula unit (Mo+2O=3 atoms)
%
% Q12. Isolated, strained MoS2 -->
%      isolated, relaxed MoS2: how much energy gain?
% MoO2/MoS2 a=2.985A mismatch: 6.1% wrt MoS2 (hexagonal lattice)
% Isolated, strained MoS2 = -21.40133632 eV
% isolated, relaxed MoS2 = -21.79870791 eV
% Energy gain = 0.39737159 eV per formula unit (Mo+2S=3 atoms)

%==================================================================

The electronic band structure of the MoO$_2$/MoS$_2$ bilayer,
shown in Figure~\ref{fig8}(a), indicates that the system is
metallic as hoped for. A superposition of electronic band
structures of isolated MoO$_2$ and MoS$_2$ monolayers, constrained
to their respective structure in the bilayer, is shown in
Figure~\ref{fig8}(b), with the Fermi levels of the individual
layers aligned. In contrast to the MoS$_2$ channel, our results
suggest that MoO$_2$ should be metallic. In view of the fact that
DFT-PBE calculations typically underestimate band gaps, we
realistically expect MoO$_2$ to be a semiconductor with a narrow,
indirect band gap instead. %
\modR{%
This is confirmed by DFT-HSE06 band structure and density of
states results, presented in the Supporting Information, which
% Fig. S3(b) and Fig. S5(b)
indicate a $0.5$~eV-wide indirect band gap in an isolated MoO$_2$
monolayer. %
}%
To inspect the level of hybridization at the MoO$_2$/MoS$_2$
interface, we superposed the band structure of the bilayer with
the bands of the MoS$_2$ monolayer, shifted rigidly up by
$0.768$~eV, and the unshifted bands of the MoO$_2$ monolayer.
Comparing the band superposition with the bilayer band structure
in Figure~\ref{fig8}(c), we find only a minimum degree of
hybridization in the system. %
\modR{%
The same low degree of hybridization, but a significant charge
transfer from MoS$_2$ to MoO$_2$ causing a net metallic behavior
of the bilayer, are also found in corresponding DFT-HSE06 band
structure results presented in the Supporting Information.
% Fig. S3(b) and Fig. S5(b)
}%

%===========< FIGURE 9 >=========================================
\begin{figure*}[t]
\includegraphics[width=1.5\columnwidth]{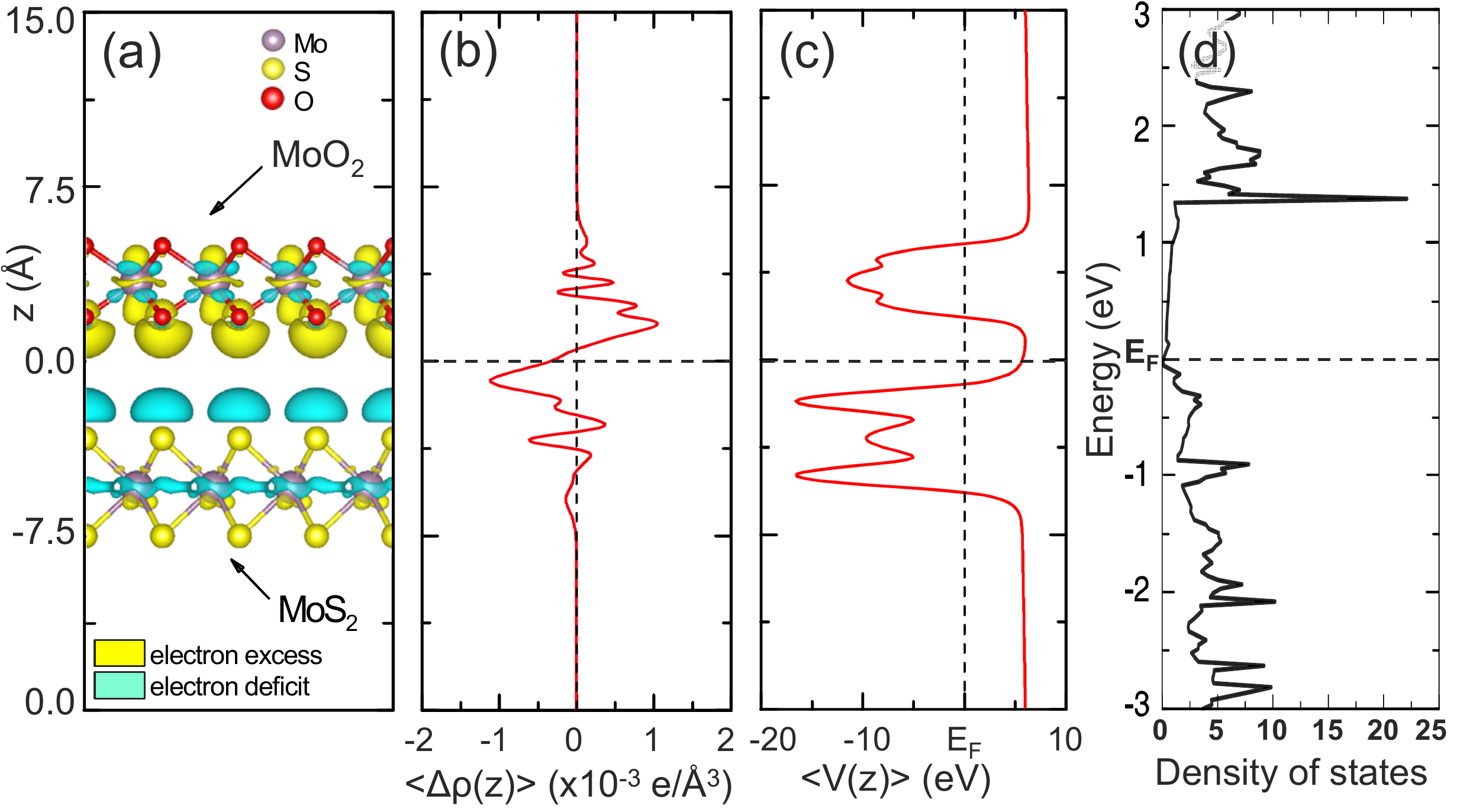}
\caption{Electronic structure changes associated with assembling
the MoO$_2$/MoS$_2$ bilayer from isolated monolayers. %
(a) Charge density difference
${\Delta}\rho=\rho({\rm{MoO}}_2/{\rm{MoS}}_2) %
-\rho({\rm{MoO}}_2) %
-\rho({\rm{MoS}}_2)$. %
${\Delta}\rho$ is shown by isosurfaces bounding regions of
electron excess at $+1.0\times10^{-3}~\text{e}$/{\AA}$^3$ (yellow)
and electron deficiency at
$-1.0\times10^{-3}~\text{e}$/{\AA}$^3$ (blue). %
(b) $\langle{\Delta}\rho(z)\rangle$ averaged across the $x-y$
plane of the layers. The black solid reference line is a guide to
the eye and $z$ indicates the position of the plane. %
(c) Electrostatic potential $<V(z)>$ in the bilayer, averaged
across the x-y plane, with $z$ denoting the position of the plane.
% The lattice constant along z direction is $30$~{\AA}.
(d) Density of states (DOS) of the bilayer, convoluted by a
Gaussian with a full-width at half-maximum of $0.1$~eV. %
%\modG{%
%Results in (a)-(c) are based on DFT-PBE. Panel (d) is the
%counterpart of DOS obtained using DFT-HSE06. %
%}%
\label{fig9} }
\end{figure*}
%===========< FIGURE 9 >=========================================

The charge redistribution in the bilayer is shown in
Figure~\ref{fig9}. The net charge transferred from the MoS$_2$ to
the MoO$_2$ layer is %
${\Delta}\rho_{2D}=1.7{\times}10^{13}$~e/cm$^2$ in the bilayer.
%${\Delta}\rho_{2D}=1.6707{\times}10^{13}$~e/cm$^2$
When divided by the thickness $t{\approx}0.4$~nm of the MoS$_2$
channel, the charge transfer density amounts to
${\Delta}\rho=4{\times}10^{20}$~e/cm$^3$, slightly lower than in
the MoO$_3$/MoS$_2$ bilayer. This doping level is still considered
to be degenerate, since $E_F$ has been moved into the valence band
of the channel. With degenerate doping in the contact region of
the bilayer, the tunnel barrier to a metal contact on the side
opposite to the doping layer should again be negligibly small and
of no consequence.

%==================================================================

\section*{Discussion}

%For the numerical computation of charge transfer, we discover that
%the charge from HSE06 method cannot directly implemented in the
%following analysis, such as Bader charge since HSE06 functional
%replaces the slowly decaying long-ranged part of the Fock
%exchange, by the corresponding density functional counterpart. The
%separation of the electron-electron interaction into a short- and
%long-ranged part, is realized only in the exchange interactions.
%Electronic correlation is represented by the corresponding part of
%the PBE density functional.\cite{{HSE03},{HSE06}} In this sense,
%the HSE method will significantly modify the charge and also
%impact the Bader result. However, we compare the charge transfer
%between nonlocal exchange-correlation (XC) PBE and local XC LDA
%methods. The results of net charge transfer $\rho_{2D}$ based on
%LDA are as following: MoO$_3$/MoS$_2$ bilayer %
%${\Delta}\rho_{2D}=5.54{\times}10^(13)$~e/cm$^2$ % MoO3/MoS2 Supercell B, LDA
%${\Delta}\rho_{2D}=4.49{\times}10^{13}$~e/cm$^2$ % MoO3/MoS2 Supercell B, PBE
%${\Delta}\rho_{2D}=2.09{\times}10^{13}$~e/cm$^2$ % MoO3/MoS2 Supercell A, LDA
%${\Delta}\rho_{2D}=1.97{\times}10^{13}$~e/cm$^2$ % MoO3/MoS2 Supercell A, PBE
%${\Delta}\rho_{2D}=1.80{\times}10^{13}$~e/cm$^2$ % MoO2/MoS2 LDA
%${\Delta}\rho_{2D}=1.67{\times}10^{13}$~e/cm$^2$ % MoO2/MoS2 PBE
%The result clearly reveals that the difference of charge transfer
%between nonlocal XC (PBE) and local XC (LDA) is small and does not
%affect us to understand the inside mechanism.

\modR{%
As expected in the outset, our numerical results based on a Bader
charge analysis indicate a significant charge transfer between
adjacent MoO$_{3-x}$ and MoS$_2$ layers in the bilayer geometry.
To validate the amount of charge transferred between the layers,
we compared the charge density difference ${\Delta}\rho_{2D}$
based on the current DFT-PBE nonlocal exchange correlation
functional to those based on the local DFT-LDA functional. The
specific values
for ${\Delta}\rho_{2D}$ in MoO$_3$/MoS$_2$ are %
$5.55{\times}10^{13}$~e/cm$^2$ (LDA for supercell B), %
% $5.5497{\times}10^(13)$~e/cm$^2$
$4.49{\times}10^{13}$~e/cm$^2$ (PBE for supercell B), %
$2.09{\times}10^{13}$~e/cm$^2$ (LDA for supercell A), %
% $2.0983{\times}10^{13}$~e/cm$^2$
$1.97{\times}10^{13}$~e/cm$^2$ (PBE for supercell A). %
The corresponding values for ${\Delta}\rho_{2D}$ in MoO$_2$/MoS$_2$ are %
$1.80{\times}10^{13}$~e/cm$^2$ (LDA) and %
% $1.8029{\times}10^{13}$~e/cm$^2$
$1.67{\times}10^{13}$~e/cm$^2$ (PBE). %
These results clearly indicate that the effect of the
exchange-correlation functional on the charge transfer is very
small. Since the treatment of exchange and correlation in the
hybrid DFT-HSE06 functional with a slowly decaying Fock exchange
is fundamentally different from the more localized form form in
DFT-LDA and DFT-PBE functionals~\cite{{HSE03},{HSE06}}, the
DFT-HSE06 functional does not provide a simple way for a Bader
charge decomposition and is not used for this purpose in our
study. %
}%

In order to eliminate the tunnel barrier between a contact metal
and a 2D semiconductor such as MoS$_2$, we have proposed a way to
locally hole dope the channel using a stable 2D material. To
achieve this objective without causing hybridization and the
emergence of mid-gap states, we chose the highly electronegative
2D semiconductor MoO$_3$ or, alternatively, MoO$_2$. Our results
confirm the expectation that the band structure in the contact
region closely resembles the superposition of energetically
shifted bands of MoS$_2$ and MoO$_3$ or MoO$_2$. We observed
degenerate hole doping of the channel in contact with
electronegative MoO$_3$ and MoO$_2$ layers. The rigid band shift
in the doped contact region has an important side effect, namely a
rigid shift of the channel bands in the doped with respect to the
undoped channel region. This is of no consequence for the vertical
tunnel barrier in the contact region, which will become negligibly
small if the channel in the conducting contact region is contacted
on the side opposite to molybdenum oxide by a metal. In this case,
we may say that we have reached the main objective of our study,
which was to optimize the metal-channel contact, using our
approach.

However, another
% a new
challenge has emerged that had been overlooked so far
to a large degree. The rigid shift of the channel bands in the
contact region causes a band offset with respect to the undoped
channel region, causing the formation of an in-plane Schottky
barrier. Since screening in a 2D system is much lower than in
conventional 3D systems, the resulting depletion region will be
larger.\cite{yi2015study} %%
%%\cite{yu2016carrier}
The in-layer Schottky barriers have to be considered seriously,
since their effect on carrier injection may be larger than that of
the vertical tunnel barriers. In that case, this lateral Schottky
barrier may dominate the contact transparency.

In principle, the band offset and the associated Schottky barrier
could be reduced by replacing the semiconducting channel material
outside the contact region by a different, isoelectronic material.
In the specific case of $p$-doped MoS$_2$, using materials with
their VBM aligned with or even higher than the shifted VBM of the
channel in the contact region will eliminate such a lateral
Schottky barrier and thus improve charge injection. Potential
candidates for in-layer contacts with the heavily $p$-doped
MoS$_2$ segment are WSe$_2$ and MoTe$_2$ as the channel
material.\cite{gong2013band} %
%% Yet creating such a perfect 2D/2D junction is a nontrivial endeavor.
Formation of such contacts by epitaxial growth of monolayer
WSe$_2$-MoS$_2$ lateral junctions with an atomically sharp
interface has recently been demonstrated.\cite{li2015epitaxial} As
an alternative, isoelectronic alloying in the channel region may
be used.\cite{duan2015synthesis} %%
Whether the Schottky barrier should be reduced by a %
lateral %%
2D/2D junction of different materials or by isoelectronic doping,
either approach will create additional interface or impurity
scattering centers.

% ==> Separate the functionality of doping and the functionality
% of a contact.

%==================================================================

\section*{Conclusions}

We have proposed an improved strategy to form low-resistance
contacts to MoS$_2$ and related semiconducting transition metal
dichalcogenides by local degenerate hole doping in the contact
region, where the atomically thin 2D channel material is in direct
contact with an electronegative material such as MoO$_3$ or
MoO$_2$. In contrast to metal contacts that are separate from the
undoped 2D channel material, the homo-junction in our study is
between undoped and heavily doped regions of the same 2D material.
To check the viability of this approach, we have determined the
equilibrium geometry and electronic structure of MoO$_3$/MoS$_2$
and MoO$_2$/MoS$_2$ bilayers and their monolayer components using
{\em ab initio} density functional calculations. Our results
indicate that, besides a rigid band shift associated with charge
transfer, the presence of molybdenum oxide modifies the electronic
structure of MoS$_2$ very little, thus avoiding the formation of
mid-gap states. We found that the charge transfer in the bilayer
provides a sufficient degree of hole doping to MoS$_2$ to render
the contact region metallic. A highly transparent contact will
thus be formed by sandwiching the semiconducting 2D MoS$_2$
channel material in-between Mo oxide and a metal.

%==================================================================

\section*{Methods}

We have studied the electronic structure, the equilibrium geometry
and structural stability of MoO$_3$ and MoO$_2$ interacting with
MoS$_2$ using {\em ab initio} density functional theory (DFT) as
implemented in the \textsc{VASP}
code~\cite{{VASP1},{VASP2},{VASP3}}. We represented these 2D
structures by a periodic array of layers separated by a vacuum
region in excess of $20$~{\AA}. %
%
%\modG{%
We used projector-augmented-wave (PAW)
pseudopotentials~\cite{{PAW1},{PAW2}}, the Perdew-Burke-Ernzerhof
(PBE)~\cite{PBE} and the {\textsc
optB86b-vdW}~\cite{{Klimes10},{Klimes11}} exchange-correlation
functionals. Since the fundamental band gap is usually
underestimated in DFT-PBE calculations, we have resorted to the
{\textsc HSE06}%
~\cite{{HSE03},{HSE06}} hybrid exchange-correlation functional, as
implemented in the {\textsc VASP}~\cite{VASP1,VASP2,VASP3,PAW2}
code, to get a different (possibly superior) description of the
band structure. We used the default mixing parameter value
$\alpha=0.25$ in these studies. %
%}%
The Brillouin zone of the conventional unit cell of the 2D
structures has been sampled by a uniform $k$-point
grid~\cite{Monkhorst-Pack76}. The specific sampling we used was
$15{\times}15$ %
for MoS$_2$, MoO$_3$ and MoO$_2$ monolayers as well as the
MoO$_2$/MoS$_2$ bilayer, %
$9{\times}9$ %
for the MoO$_3$/MoS$_2$ bilayer with the supercell A containing 14
atoms, and
$2{\times}2$ %
for the MoO$_3$/MoS$_2$ bilayer with supercell B containing 156
atoms.
%MoO3/MoS2 Supercell A: 9x9 %
%MoO3/MoS2 Supercell B: 2x2 %
%MoO2/MoS2: 15x15 %
%MoO3 (ML): 15x15 %
%MoO2 (ML): 15x15 %
%MoS2 (ML): 15x15 %
We used $550$~eV as the electronic kinetic energy cutoff for the
plane-wave basis and a total energy difference between subsequent
self-consistency iterations below $10^{-5}$~eV/atom
% for entire unit cell
% 14 atoms supercell A
% 156 atoms supercell B $10^{-3}$~eV <10-5 eV/atom
as the criterion for reaching self-consistency. All geometries
have been optimized using the conjugate-gradient
method~\cite{CGmethod}, until none of the residual
Hellmann-Feynman forces exceeded $10^{-3}$~eV/{\AA}. The maximum
force criterion has been relaxed to $10^{-2}$~eV/{\AA} in the
optimization of the MoO$_3$/MoS$_2$ bilayer with the large
supercell B containing 156 atoms.

%==================================================================

\begin{suppinfo}
\modR{%
Supercells used to represent the incommensurate MoO$_3$/MoS$_2$
bilayer. Comparison between the electronic band structure and
density of states of MoO$_3$/MoS$_2$ and MoO$_2$/MoS$_2$ obtained
using the DFT-HSE06 and DFT-PBE exchange-correlation functionals. %
}%
\end{suppinfo}

%%%%%%%%%%%%%%%%%%%%%%%%%%%%%%%%%%%%%%%%%%%%%%%%%%%%%%%%%%%%%%%%%%%%%
% Author information
%%%%%%%%%%%%%%%%%%%%%%%%%%%%%%%%%%%%%%%%%%%%%%%%%%%%%%%%%%%%%%%%%%%%%

\quad\\
{\noindent\bf Author Information}\\

{\noindent\bf Corresponding Author}\\
$^*$E-mail: {\tt tomanek@pa.msu.edu} \\

% {\noindent\bf Notes}\\
% The authors declare no competing financial interest.

%%%%%%%%%%%%%%%%%%%%%%%%%%%%%%%%%%%%%%%%%%%%%%%%%%%%%%%%%%%%%%%%%%%%%
%% Acknowledgements should use the acknowledgement environment.
%%%%%%%%%%%%%%%%%%%%%%%%%%%%%%%%%%%%%%%%%%%%%%%%%%%%%%%%%%%%%%%%%%%%%

\begin{acknowledgement}
D.T. acknowledges partial support by the NSF/AFOSR EFRI 2-DARE
grant number \#EFMA-1433459. Z.G. acknowledges financial support
from the China Scholarship Council under Grant No. 201706260027
and the hospitality of Michigan State University. Z.Z.
acknowledges partial support by NSF grant number DMR-1308436 and
the WSU Presidential Research Enhancement Award. Z.G. acknowledges
useful discussions with Jie Ren. Computational resources have been
provided by the Michigan State University High Performance
Computing Center.
\end{acknowledgement}

%\end{linenumbers}

% \bibliography{MoOxTMD19}
% \end{document}
%%%%%%%%%%%%%%%%%%%%%%%%%%%%%%%%%%%%%%%%%%%%%%%%%%%%%%%%%%%%%%%%%%%%%

\providecommand{\latin}[1]{#1} \makeatletter \providecommand{\doi}
  {\begingroup\let\do\@makeother\dospecials
  \catcode`\{=1 \catcode`\}=2 \doi@aux}
\providecommand{\doi@aux}[1]{\endgroup\texttt{#1}} \makeatother
\providecommand*\mcitethebibliography{\thebibliography} \csname
@ifundefined\endcsname{endmcitethebibliography}
  {\let\endmcitethebibliography\endthebibliography}{}

\end{document}